\documentclass[pra,twocolumn,showpacs,floatfix]{revtex4}
\usepackage[dvips]{graphicx}
\usepackage{hyperref}
\usepackage{bm}
\usepackage{dsfont}
\usepackage{rotating}
\usepackage{times}

\begin{document}
\def\diff{\mathrm{d}}

\def\jpb{J. Phys. B: At. Mol. Opt. Phys.~}
\def\pra{Phys. Rev. A~}
\def\prb{Phys. Rev. B~}
\def\prl{Phys. Rev. Lett.~}
\def\jmo{J. Mod. Opt.~}
\def\jetp{Sov. Phys. JETP~}
\def\etal{{\em et al.}}

\def\pabl#1#2{\frac{\partial #1}{\partial #2}}
\def\abl#1#2{\frac{\diff #1}{\diff #2}}

\def\varphit{\tilde{\varphi}}
\def\Mtilde{\tilde{M}}
\def\Ktilde{\tilde{K}}
\def\gammatilde{\tilde{\gamma}}

\def\twovec#1#2{\left( \begin{array}{c} #1 \\ #2 \end{array}\right)}
\def\twotimestwo#1#2#3#4{\left( \begin{array}{cc} #1 & #2 \\ #3 & #4 \end{array}\right)}

\def\vekt#1{{\bm{#1}}}
\def\vect#1{\vekt{#1}}
\def\vektalpha{\vekt{\alpha}}
\def\vektr{\vekt{r}}
\def\vektd{\vekt{d}}
\def\vektx{\vekt{x}}
\def\vektp{\vekt{p}}
\def\NRNOmax{N_{\mathrm{RNO}}}
\def\Ncyc{N_{\mathrm{cyc}}}
\def\Vn{V_{\mathrm{n}}}
\def\Vi{V_{\mathrm{i}}}
\def\Hi{H_{\mathrm{i}}}
\def\vee{v_{\mathrm{ee}}}
\def\ee{{\mathrm{ee}}}
\def\fee{f_{\mathrm{ee}}}
\def\vektpc{\vekt{p}_{\mathrm{c}}}
\def\vektE{\vekt{E}}
\def\vekte{\vekt{e}}
\def\vektA{\vekt{A}}
\def\vektEhat{\hat{\vekt{E}}}
\def\vektB{\vekt{B}}
\def\vektv{\vekt{v}}
\def\vektk{\vekt{k}}
\def\vektkhat{\hat{\vekt{k}}}
\def\reff#1{(\ref{#1})}

\def\calH{{\cal H}}
\def\calD{{\cal D}}
\def\calh{{\cal h}}
\def\calM{{\cal M}}
\def\calK{{\cal K}}

\def\Tr{{\mathrm{Tr}}}
\def\Up{U_{\mathrm{p}}}
\def\Ip{I_{\mathrm{p}}}
\def\Tp{T_{\mathrm{p}}}
\def\Ss{S_{\vektp\mathrm{s}}}
\def\SIs{S_{\vektp\Ip\mathrm{s}}}
\def\Cps{C_{\vektp\mathrm{s}}}
\def\Cpnulls{C_{\vektp_0\mathrm{s}}}
\def\SI{S_{\vektp\Ip}}
\def\SIsnull{S_{\vektp_0\Ip\mathrm{s}}}
\def\zt{z_{\mathrm{t}}}
\def\ts{t_{\vektp\mathrm{s}}}
\def\tsnull{t_{\vektp_0\mathrm{s}}}
\def\tnulltilde{\tilde{t}_0}
\def\omegap{\omega_p}
\def\omegaMie{\omega_M}
\def\imagi{\mathrm{i}}
\def\eulere{\mathrm{e}}

\def\halb{\frac{1}{2}}

\def\phitilde{\tilde{\phi}}
\def\alphatilde{\tilde{\alpha}}
\def\phitildeperp{\tilde{\phi}^{\perp}}

 \def\psihat{{\hat{\psi}}}
 \def\psihatdag{{\hat{\psi}^\dagger}}
\def\beq{\begin{equation}}
\def\eeq{\end{equation}}

\def\energy{{\cal E}}

\def\Ehat{\hat{E}}
\def\Ahat{\hat{A}}
\def\ahat{\hat{a}}
\def\ahatdag{\hat{a}^\dagger}

\def\ket#1{\vert #1\rangle}
\def\bra#1{\langle#1\vert}
\def\braket#1#2{\langle #1 \vert #2 \rangle}

\def\makered#1{{\color{red} #1}}

\def\Re{\,\mathrm{Re}\,}
\def\Im{\,\mathrm{Im}\,}

\def\varphic{\varphi_{\mathrm{c}}}

\def\tini{t_\mathrm{i}}
\def\tfinal{t_\mathrm{f}}
\def\vxc{v_\mathrm{xc}}
\def\vextop{\hat{v}_\mathrm{ext}}
\def\VC{V_\mathrm{c}}
\def\VHX{V_\mathrm{Hx}}
\def\VHXC{V_\mathrm{Hxc}}
\def\wop{\hat{w}}
\def\Gammaevenodd{\Gamma^\mathrm{even,odd}}
\def\Gammaeven{\Gamma^\mathrm{even}}
\def\Gammaodd{\Gamma^\mathrm{odd}}

\def\Hop{\hat{H}}
\def\hop{\hat{h}}
\def\Cop{\hat{C}}
\def\HopKS{\hat{H}_\mathrm{KS}}
\def\HKS{H_\mathrm{KS}}
\def\Top{\hat{T}}
\def\TopKS{\hat{T}_\mathrm{KS}}
\def\VopKS{\hat{V}_\mathrm{KS}}
\def\VKS{{V}_\mathrm{KS}}
\def\vKS{{v}_\mathrm{KS}}
\def\Ttildeop{\hat{\tilde{T}}}
\def\Ttilde{{\tilde{T}}}
\def\Vextop{\hat{V}_{\mathrm{ext}}}
\def\Vext{V_{\mathrm{ext}}}
\def\Vopee{\hat{V}_{{ee}}}
\def\psiopdag{\hat{\psi}^{\dagger}}
\def\psiop{\hat{\psi}}
\def\vext{v_{\mathrm{ext}}}
\def\Vee{V_{ee}}
\def\nop{\hat{n}}
\def\Uop{\hat{U}}
\def\Wop{\hat{W}}
\def\bop{\hat{b}}
\def\bopdag{\hat{b}^{\dagger}}
\def\qop{\hat{q}}
\def\jop{\hat{j\,}}
\def\vHxc{v_{\mathrm{Hxc}}}
\def\vHx{v_{\mathrm{Hx}}}
\def\vH{v_{\mathrm{H}}}
\def\vc{v_{\mathrm{c}}}
\def\xop{\hat{x}}

\def\Wcmcm{W/cm$^2$}

\def\varphiexact{\varphi_{\mathrm{exact}}}

\def\fmathbox#1{\fbox{$\displaystyle #1$}}

\title{Time-dependent renormalized natural orbital theory applied to the two-electron spin-singlet case: ground state, linear response, and autoionization}

\author{M.~Brics}
\affiliation{Institut f\"ur Physik, Universit\"at Rostock, 18051 Rostock, Germany}

\author{D.~Bauer}
\thanks{Corresponding author: dieter.bauer@uni-rostock.de}
\affiliation{Institut f\"ur Physik, Universit\"at Rostock, 18051 Rostock, Germany}

\date{\today}

\begin{abstract} 
Favorably scaling numerical time-dependent many-electron techniques such as time-dependent density functional theory (TDDFT) with adiabatic exchange-correlation potentials typically fail in capturing highly correlated electron dynamics. We propose a method based on natural orbitals, i.e., the eigenfunctions of the one-body reduced density matrix, that is almost as inexpensive numerically as adiabatic TDDFT, but which is capable of describing correlated phenomena such as doubly excited states, autoionization, Fano profiles in the photoelectron spectra, and strong-field ionization in general. Equations of motion (EOM) for natural orbitals and their occupation numbers have been derived earlier. 
We show that by using renormalized natural orbitals (RNO) both can be combined into one equation governed by a hermitian effective Hamiltonian. We specialize on the two-electron spin-singlet system, known as being a ``worst case'' testing ground for TDDFT, and employ the widely used, numerically exactly solvable, one-dimensional helium model atom (in a laser field) to benchmark our approach.    The solution of the full, nonlinear EOM for the RNO  is plagued by instabilities, and resorting to linear response is not an option for the ultimate goal to study nonperturbative dynamics in intense laser fields. We therefore make two rather bold approximations: we employ the initial-state-``frozen'' effective RNO Hamiltonian for the time-propagation and truncate the number of RNO to only two per spin. Surprisingly, it turns out that even with these strong approximations we obtain a highly accurate ground state, reproduce doubly-excited states, and autoionization.

\end{abstract}
\maketitle


\section{Introduction}
The ``holy grail'' of computational many-body theory is to overcome the so-called ``exponential wall,'' i.e., the exponentially increasing numerical effort as a function of the particle number to solve the many-body Schr\"odinger equation \cite{Kohn99}. It is an obvious idea that one should try to replace the high-dimensional many-body wavefunction by some simpler, lower-dimensional quantity, and then derive equations governing this quantity because the ``many-electron wavefunction tells us more than we need to know'' \cite{coulson}. Reduced density matrices (RDM) appear to be most suitable for that purpose. In fact, extensive research has been devoted to the properties and applications of RDM, starting with the classic work by L\"owdin \cite{loewdin1}, already early summarized in \cite{colemanrmp}, and meanwhile covered in books 
and reviews \cite{colemanyukalov,cios,gido,mazz,mazz2}, and excellent thesis works \cite{appelthesis,giesbthesis}. 

For systems with two-body interactions, any observable can be {\em explicitly} written down in terms of the  two-body reduced density matrix (2RDM). The Hohenberg-Kohn theorem \cite{HohenbergKohn} of density functional theory (DFT) (see, e.g., \cite{dreizler}) even ensures that any observable (of a system governed by a Hamiltonian with a scalar, local, external potential) is {\em in principle} a functional of the single-particle density, i.e., the diagonal of the (spin-integrated) one-body reduced density matrix (1RDM). However, these functionals are not known for all observables of interest so that approximations have to be made in practice. Quite reasonably, it seems that the more reduced the quantity employed is, the more approximations have to be made in the governing equations (such as the intricate exchange-correlation (xc) potential in the Kohn-Sham scheme \cite{KohnSham}), and for the observables. Employing the 1RDM as the basic ``variable'' instead of the single-particle density brings us to reduced density matrix functional theory (RDMFT) \cite{loewdin1,colemanrmp,colemanyukalov,cios,gido,mazz,appelthesis,giesbthesis}.  In RDMFT, the Bogoliubov-Born-Green-Kirkwood-Yvon  chain of equations needs to be truncated by a (sufficiently accurate) approximation of the 2RDM as a functional of the 1RDM. In the simplest form, this leads just to the Hartree-Fock (HF) equations. Expressions for the 2RDM  beyond HF have been devised and applied (see, e.g., \cite{mueller,goeumrig,grit,piris,zarkadoula}), although  relatively few compared to the abundant literature on xc potentials in DFT.
Approaches using directly the 2RDM as the basic variable have been proposed and applied as well \cite{mazz2}. A naive minimization of the energy as a functional of the 2RDM will, however, yield too low energy values, as not all two-body density matrices (2DM) originate from an $N$-electron state. Recent progress in the solution of this so-called $N$-representability problem has been made by formulating a hierarchy of conditions the 2DM has to fulfill (without resorting to higher-order RDM) \cite{mazziottiPRL2012}.

It is computationally beneficial to expand the 1RDM and 2RDM in 1RDM eigenfunctions, the so-called natural orbitals (NO) \cite{loewdin1}, as they form the best possible basis set (in a well-defined mathematical sense; see, e.g., \cite{colemanyukalov,giesbthesis} for details). The resulting equations for these NO form a set of coupled, nonlinear Schr\"odinger-like equations \cite{pernal}, as in configuration interaction calculations. The eigenvalue of the 1RDM to which  a NO belongs can be interpreted as its occupation number (OCN). Unlike in, e.g., Hartree-Fock, these OCN are, in general, fractional $\in ]0,1[$ in correlated fermionic systems (unless they are ``pinned'' to $0$ or $1$ \cite{helbigpinned,schillingpinned}).

In this work, we investigate whether NO can be efficiently employed to describe the correlated {\em dynamics} of a two-electron spin-singlet system in an external, driving field such as that of a laser. Of course, the study of the {\em structure} of correlated two-electron systems has a long history that started soon after the ``invention'' of quantum mechanics. In view of a NO description it has been analyzed by L\"owdin and Shull in 1956 \cite{loewdintwoelecs}. The two-electron spin-singlet ground state wavefunction has an exceptionally simple structure when expanded in NO, as the coefficient matrix turns out to be diagonal. This means that the ground state 2DM   needs not to be approximated in terms of the NO but is known exactly  in the two-electron case.  

One may think that the two-electron case is a bit (too) trivial to test a novel time-dependent {\em many}-body method. However, this is not the case. While electronic structure calculations are difficult enough, time-dependent quantum dynamics beyond linear response with non-perturbative drivers is by far more challenging. In fact, from a computational point of view it is orders of magnitude more demanding because on top of the ground state problem one, subsequently, needs to propagate the system for typically $10^3$--$10^4$  time steps on numerical grids typically a $10^2$--$10^3$ times larger than that for the ground state. It is thus even more important to develop efficient numerical methods capable of describing such strongly driven quantum dynamics. Time-dependent density functional theory (TDDFT) \cite{ullrich,marques} works well in many cases but fails (with known and practicable adiabatic xc-potentials) whenever the processes to be described rely on strong correlation or involve resonant interaction \cite{leinkuemmel,wilken,wilken2,thiele,wijn,ruggbau,helbig,raghunest,elliott}. As the correlation energy (relative to the total energy) typically increases as the number of electrons decreases (down to two), it is the {\em few}-body correlated electron dynamics that serve as ``worst case'' benchmarks for methods beyond TDDFT with adiabatic xc potentials. For instance, autoionization in strong laser fields is currently investigated experimentally \cite{wang,ott} and particularly challenging for theory because it involves multiply-excited states. As multiply excited states are absent in TDDFT using adiabatic xc potentials   \cite{ullrich,maitraautoioni} it serves as an ideal testing ground for novel {\em ab initio} methods going beyond ``standard'' TDDFT.

An algorithm for propagating NO and OCN for two coupled nonlinear oscillators has been proposed in  \cite{portes}. The general equations of motion (EOM) for the NO and their OCN have been derived in \cite{pernalTD} (see also \cite{appelthesis}). However, the volume of published work on {\em time-dependent} density matrix functional theory (TDDMFT) is still very limited. Different adiabatic approximations to TDDMFT have been derived and applied to molecules \cite{pernalgiesbertz,gies1,gies2,gies3} and a two-site Hubbard model \cite{requist}, respectively. Exact time-dependent NO occupations have been investigated \cite{appelgross} using the same numerically exactly solvable model atom employed in the current work. It was found that common approximations for the 2RDM functional render the  OCN constant, which is incorrect for, e.g.,  atoms in strong laser pulses or resonant interactions \cite{helbig}. A semi-classical approach to propagate the 1RDM that allows for changing OCN has been proposed in \cite{rajam} and applied to Moshinsky's two-electron model atom \cite{moshinsky}.

Our paper is organized as follows. In Sec.~\ref{sec:theo} we review the basic density matrix and NO theory for the two-electron case and introduce renormalized NO (RNO), which allow to unify the EOM for the OCN and the NO. Further, we specialize on the spin-singlet case, briefly discuss the time-dependent Hartree-Fock limit, and derive variationally the equations governing the RNO ground state configuration. In Sec.~\ref{sec:resanddisc} we present results for two RNO per spin. After the ground state is obtained, the linear response spectrum and autoionization in a laser field are investigated. We conclude and give an outlook in Sec.~\ref{sec:concl}.  Details of the derivation of the EOM for the NO are given in Appendix \ref{sec:derivationI}, the relation between the expansion coefficients of the ground state 2DM and those of the two-electron spin singlet ground state is given in Appendix \ref{app:gamma2andD}.

Atomic units are used throughout.

\section{Theory} \label{sec:theo}
 The hermitian 2DM for a two-electron system with wavefunction $\Phi(12;t)$ reads 
\beq \gamma_2(12,1'2';t)=\Phi^*(1'2';t) \Phi(12;t),\label{gamma2} \eeq 
where the arguments $1$, $2$, $1'$ etc.\ comprise spatial and spin-degrees of freedom $(x_1,\sigma_1)$, $(x_2,\sigma_2)$, $(x_1',\sigma_1') \ldots$.   
The hermitian 1RDM is
\beq \gamma_1(1,1';t)=2 \int\diff 2\, \gamma_2(12,1'2;t).  \label{gamm1andgamma2} \eeq

Given a Hamiltonian for the two-electron time-dependent Schr\"odinger equation (TDSE) 
\beq \imagi\partial_t \Phi(12;t) =  \hat H(12;t)\Phi(12;t) \eeq
of the form
\beq \hat H(12;t) = \hat h_0(1;t) + \hat h_0(2;t) + \vee(12), \eeq
where $\vee$ is the electron-electron interaction,
 the  2DM fulfills the von Neumann equation while the 1RDM obeys the EOM
\begin{eqnarray}
\lefteqn{-\imagi\partial_t \gamma_1(1,1';t) =  [\hat h_0(1') -  \hat h_0(1)] \gamma_1(1,1';t)} \nonumber \\ && + 2 \int \diff 2 \,  \bigl\{ \vee(1'2) - \vee(12) \bigr\} \gamma_2(12,1'2;t). \label{eom1rdm}
\end{eqnarray}

The hermitian 1RDM can be written in terms of an orthonormalized set of NO $\phi_k$, $k=1,2,3, \ldots$, and real, positive-definite OCN $n_k$ as
\beq \gamma_1(1,1';t) = \sum_{k} n_k(t) \phi_k^*(1';t) \phi_k(1;t), \label{gamma1rdm}\eeq
\beq  \sum_k n_k(t)= 2, \quad \int\diff 1\, |\phi_k(1;t)|^2=1 . \label{ocsandnorm} \eeq
In other words, $\phi_k$ is the eigenvector of $\gamma_1$ with respect to the eigenvalue $n_k$,
\beq \int\diff 1'\, \gamma_1(1,1';t)  \phi_k(1';t) = n_k(t)\phi_k(1;t) . \eeq
As the NO form a complete basis one can expand the 2DM in them, 
\begin{eqnarray}\lefteqn{ \gamma_2(12,1'2';t)}  \label{gamma2rdm}\\
& =& \sum_{ijkl} \gamma_{2,ijkl}(t) \phi_i(1;t) \phi_j(2;t) \phi_k^*(1';t) \phi_l^*(2';t).  \nonumber \end{eqnarray}

\subsection{Equation of motion for renormalized natural orbitals}\label{sec:eomrno}
We find it numerically beneficial to incorporate the OCN into the NO by renormalizing them,
\beq \phitilde_k(1;t)={\sqrt{n_k(t)}}\ \phi_k(1;t), \label{phitildes} \eeq
\beq \int\diff 1\, | \phitilde_k(1;t)|^2 = n_k(t). \label{normalizedphis} \eeq
Inserting  the NO-expansions  of the density matrices $\gamma_1$ and $\gamma_2$ \reff{gamma1rdm} and \reff{gamma2rdm}, respectively, into the EOM for the 1RDM \reff{eom1rdm},  coupled EOM for the NO and OCN have been derived and reviewed in the literature \cite{pernalTD,appelthesis,giesbthesis}. 

We have derived a single EOM that is particularly useful for our purposes. It  has the form
\beq \imagi\partial_t \tilde{ \boldsymbol \Phi}(1;t) = \hat \calH(1;t) \tilde{ \boldsymbol\Phi}(1;t)\label{eq:desiredform} \eeq
with a hermitian Hamiltonian $ \hat \calH$ and a column vector $\tilde{ \boldsymbol\Phi}(1;t)$ with the RNO $\phitilde_k(1;t)$ in it. The derivation is given in Appendix \ref{sec:derivationI}, the result being
\begin{widetext}
\begin{eqnarray}
\imagi \partial_t \phitilde_n(1;t) &=& -\frac{1}{n_n(t)}   \left\{ 2   \Re  \sum_{jkl} \tilde \gamma_{2,njkl}(t) \bra{\tilde k(t)\tilde l(t)}  \vee \ket{\tilde n(t)\tilde j(t)}+  \bra{\tilde n(t)}\hat h_0(t) \ket{\tilde n(t)}\right\}\phitilde_n(1;t) \nonumber\\
&& + \sum_{k\neq n}  \frac{2}{n_k(t)-n_n(t)}  \sum_{jpl} \left\{\tilde \gamma_{2,kjpl}(t) \bra{\tilde p(t)\tilde l(t)}  \vee \ket{\tilde n(t)\tilde j(t)}   -  \bigl[\tilde \gamma_{2,njpl}(t)  \bra{\tilde p(t)\tilde l(t)} \vee \ket{\tilde k(t)\tilde j(t)}   \bigr]^*  \right\} \phitilde_k(1;t) \nonumber \\
&& + \hat h_0(1;t) \phitilde_n(1;t) + 2 \sum_k  \sum_{jl} \tilde\gamma_{2,kjnl}(t) \bra{\tilde l(t)} \vee \ket{\tilde j(t)}(1;t) \, \phitilde_k(1;t).
\label{eomrenormphis} \end{eqnarray}
\end{widetext}
Here,  we used the abbreviations
\beq \gammatilde_{2,njkl}(t)=\frac{\gamma_{2,njkl}(t)}{\sqrt{n_n(t)n_j(t)n_k(t)n_l(t)}},\label{eq:gammatilde}\eeq
\beq \bra{\tilde l(t)} \vee \ket{\tilde j(t)}(1;t) = \int\diff 1'\, \phitilde^*_l(1';t) \vee(11') \phitilde_j(1';t), \label{matrixelement1} \eeq
\begin{eqnarray} \lefteqn{\bra{\tilde k(t)\tilde l(t)}  \vee \ket{\tilde n(t)\tilde j(t)}}\nonumber \\
&&  = \int\diff 1\,  \phitilde^*_k(1;t) \bra{\tilde l(t)} \vee \ket{\tilde j(t)}(1;t)  \phitilde_n(1;t).\label{matrixelement2} \end{eqnarray}
With  \reff{normalizedphis}, Eq.~\reff{eomrenormphis} is a set of coupled  EOM for the RNO alone. 

Multiplication of \reff{eomrenormphis} by $\tilde \phi_n^*(1;t)$ and integration $\int\diff 1$ yields
\beq \dot n_n(t) = - 4 \,\Im \sum_{jkl} \tilde \gamma_{2,njkl}(t) \bra{\tilde k(t)\tilde l(t)} \vee \ket{\tilde n(t)\tilde j(t)} ,\label{ndotn} \eeq
which, expressed in terms of NO  instead of RNO, has been derived earlier \cite{pernalTD}.
This equation is useful to see whether a certain approximation of $\tilde \gamma_{2,njkl}$ will lead to time-varying OCN or constant OCN. For instance, common approximations of the form
\beq \tilde \gamma_{2,njkl}^{\mathrm{(approx)}}(t)=f_{njkl}(t) \delta_{nk}\delta_{jl} - g_{njkl}(t) \delta_{nl}\delta_{jk} \label{gamma2approx}\eeq
with $f_{njkl}(t)$ and  $g_{njkl}(t)$ real will lead to constant OCN, $ \dot n_n(t) \equiv 0$, because all NO phases that could lead to an imaginary part on the right hand side of \reff{ndotn} cancel.  From the exact numerical solution of the two-electron TDSE we know that for, e.g., resonant interactions (Rabi floppings) and in other scenarios the OCN {\em do} change in time \cite{helbig,appelgross,rajam}.

If the total number of particles is conserved,
  \beq 0= \Im \sum_{njkl} \tilde \gamma_{2,njkl}(t) \bra{\tilde k(t)\tilde l(t)} \vee \ket{\tilde n(t)\tilde j(t)} \label{ndotn2} \eeq
follows from \reff{ndotn} upon summing over $n$. Moreover, one finds that the RNO stay mutually orthogonal if 
\beq  \calD_{nm}(t)=\calD_{mn}^*(t)\label{eq:Dnm} \eeq
holds, where
\beq\calD_{nm}(t)= \sum_{kjl}\tilde\gamma_{2,kjnl}(t) \bra{\tilde m(t) \tilde l(t)} \vee \ket{\tilde k(t)\tilde j(t)}. \label{Dhermitian}\eeq 
As the RNO, being $\forall t$ eigenfunctions of a hermitian matrix, {\em should} stay orthogonal, Eq.~\reff{eq:Dnm}
 poses a condition any approximation of $\tilde\gamma_{2,kjnl}(t)$ has to fulfill.

\subsection{Two-electron spin-singlet case}\label{sec:twoelspsing}
If the two-electron wavefunction is $\forall t$  of the spin-singlet form
\beq \Phi(12;t) = \Phi(x_1x_2;t) \frac{1}{\sqrt{2}} (\delta_{\sigma_1+}\delta_{\sigma_2-} - \delta_{\sigma_1-}\delta_{\sigma_2+} ), \label{twoelspinsingletwf}\eeq $ \Phi(x_1x_2;t) =  \Phi(x_2x_1;t)$, (with ``$+$,'' ``$-$'' for ``spin-up'' and ``spin-down,'' respectively),
we have with \reff{gamma2} and
\beq \gamma_1(x_1,x_1';t) = 2 \int\diff x_2\,\Phi^*(x_1'x_2;t) \Phi(x_1x_2;t) \label{gamma1x1x1prime} \eeq 
that
\beq
\gamma_1(1,1';t) = \halb \gamma_1(x_1,x_1';t) (\delta_{\sigma_1+}\delta_{\sigma_1'+} + \delta_{\sigma_1-}\delta_{\sigma_1'-} ).\label{gamma1spinsinglet} \eeq

We switch temporarily back to the NO normalized to unity. Making the ansatz 
\beq \phi_k(x\sigma;t) = \phi_k(x;t) \, (a_k\delta_{\sigma +} + b_k \delta_{\sigma -}),\eeq
\beq  |a_k|^2 + |b_k|^2= 1, \quad \int\diff x\, |\phi_k(x;t)|^2 = 1  \eeq
one simply finds the same form for the spin-reduced 1RDM as in   \reff{gamma1rdm},
\beq \gamma_1(x,x';t) = \sum_k n_k(t) \phi_k^*(x';t) \phi_k(x;t) .\label{gamma1xxprimeinNO} \eeq
Because of orthonormality of the NO, for $k\neq k'$
\beq (a^*_{k'} a_k + b^*_{k'}b_k)\int\diff x \,  \phi_{k'}^*(x;t)\phi_k(x;t)  = 0\eeq
need to be fulfilled
so that either the spatial part of the NO must be orthogonal or the spin-part. For a given spin-part of $\phi_k$ with coefficients $a_k$ and $b_k$ one can always find
(up to an irrelevant phase factor) a normalized spin-part $(a_{k'}\delta_{\sigma +} + b_{k'}\delta{\sigma_-})$ of $\phi_{k'}$  that is orthogonal to  $(a_{k}\delta_{\sigma +} + b_{k}\delta_{\sigma_-})$.
A convenient choice is $ a_k=1$, $ b_k=0$, $a_{k'}=0$, $b_{k'}=1$ 
while $\phi_{k'}(x;t) = \phi_{k}(x;t)$. 
In other words, the spatial NO appear {\em pairwise} equal, with opposite spin parts. This is probably the most trivial example of a {\em pairing} phenomenon. 
 One may order the NO such that for odd $k$ the following equations hold:
\beq \phi_k(x\sigma;t)= \phi_k(x;t)\,\delta_{\sigma +}, \qquad k=1,3,5,  \ldots,\eeq
\beq \phi_k(x;t)=\phi_{k+1}(x;t),\eeq
\beq \phi_{k+1}(x\sigma;t) = \phi_{k+1}(x;t)\,\delta_{\sigma -}= \phi_{k}(x;t)\,\delta_{\sigma -} . \label{ordering}\eeq
Then we can write instead of \reff{gamma1xxprimeinNO}
\beq \gamma_1(x,x';t)= 2 \sum_{k\ \mathrm{odd}} n_k(t) \phi_k^*(x';t) \phi_k(x;t), \label{gamma1odd}\eeq
and we need to consider only half of the NO (i.e., those with, e.g., the odd indices, i.e., ``spin-up'') in the following.

Because the spatial spin-singlet wavefunction is symmetric it can be shown \cite{giesbthesis} that its expansion in NO is diagonal,
\beq  \Phi(x_1x_2;t) = \sum_{i \ \mathrm{odd}}  D_i(t)\, \phi_i(x_1;t)\phi_i(x_2;t), \label{expandphix1x2II}
\eeq
where, because of \reff{gamma1x1x1prime} and \reff{gamma1xxprimeinNO},
\beq |D_i(t)|^2 = n_i(t) .\label{Dimodsqu} \eeq
We can thus use the NO expansion coefficients for the wavefunction $D_i(t)$ instead of resorting to the NO expansion coefficients $\gamma_{2,njkl}(t)$ for the 2DM. How both are connected is discussed in Appendix~\ref{app:gamma2andD}. 
Formally, in Eq.~\reff{expandphix1x2II} the time-dependent spatial two-electron wavefunction is written as a single {\em geminal} \cite{mazziottiJChemPhys2000}, expanded in time-dependent NO.

The EOM for the spatial RNO in terms of the time-dependent geminal expansion coefficients $D_i$ can be written as (from now on all time-arguments suppressed for brevity)
\beq 
\imagi \partial_t \phitilde_n(x) = \hat\calH^0_n(x) \phitilde_n(x) + \sum_{k\ \mathrm{odd}\ \neq n}  \calH^1_{nk}(x) \phitilde_k(x), \label{tdphitilde}
\eeq
where
\begin{eqnarray}
\hat\calH^0_n(x) &=& A_n + \hat\calK^0_n(x),\label{H0n} \\
\calH^1_{nk}(x) &=& B_{nk} + \calK^1_{nk}(x) 
\end{eqnarray}
with
\begin{widetext}
\begin{eqnarray}
A_n &=&  - \frac{1}{n_n} \left[   \Re\left(   \sum_{k\ \mathrm{odd}} \frac{D_n\, D_k^*}{n_n n_k}
 \bra{\tilde k\tilde k}  \vee \ket{\tilde n \tilde n}_x\right)
+ \bra{\tilde n}\hat h_0 \ket{\tilde n}_x \right], \quad \hat\calK^0_n(x) =  \hat h_0(x) +  
\frac{\bra{\tilde n} \vee \ket{\tilde n}_x(x)}{n_n},  \\
B_{nk} &=&  \frac{1}{n_k-n_n}  \sum_{p\ \mathrm{odd}} \left( \frac{D_k\, D_p^*}{n_k n_p}
\bra{\tilde p\tilde p}  \vee \ket{\tilde n\tilde k}_x -  \left[
\frac{D_n\, D_p^*}{n_nn_p}
\bra{\tilde p\tilde p} \vee \ket{\tilde k\tilde n}_x    \right]^*  \right), \quad \calK^1_{nk}(x)  = \frac{D_k\, D_n^*}{n_kn_n}
\bra{\tilde n} \vee \ket{\tilde k}_x(x),\label{k1nk}
\end{eqnarray}
\end{widetext}
and the potentials $\bra{\tilde n} \vee \ket{\tilde k}_x(x)$ and matrix elements $\bra{\tilde p\tilde p}  \vee \ket{\tilde n\tilde k}_x$ defined as in \reff{matrixelement1} and \reff{matrixelement2} but all integrals with respect to position space only. Here we exploit that the electron-electron interaction does not directly involve the spin degrees of freedom. 


\subsection{Time-dependent Hartree-Fock limit} \label{sec:tdhflimit}
Equations \reff{tdphitilde}--\reff{k1nk} reduce to the two-electron spin-singlet Hartree-Fock limit for $n_1=n_2=1$, $\tilde\phi_1(x)=\phi_1(x)=\tilde\phi_2(x)=\phi_2(x)$, and all other OCN (and thus RNO) zero. The (in the NO-index) off-diagonal part of the Hamiltonian $\calH^1_{nk}(x)$ in \reff{tdphitilde} therefore vanishes, and
\[ A_1=-\bra{ 1 1} \vee \ket{ 1 1}_x-\bra{ 1} \hat h_0 \ket{ 1}_x, \]
\[ \hat\calK^0_1(x) =  \hat h_0(x) +   \bra{1} \vee \ket{1}_x(x) \]
so that
\[ \imagi\partial_t \phi_1(x) =  [ \hat h_0(x) +   \bra{1} \vee \ket{1}_x(x) +A_1 ]\phi_1(x) , \]
which is indeed the time-dependent Hartree-Fock (TDHF) equation for the two-electron spin-singlet system (where the Fock-term cancels half of the Hartree). In the case of the two-electron spin-singlet system TDHF is equivalent to a time-dependent Kohn-Sham (TDKS) treatment in ``exact-exchange approximation'' (EXA) and the correlation potential set to zero.  Note that, unlike in the general case with more than one NO per spin, the purely time-dependent term $A_1$ can be eliminated via a contact transformation $\phi_1(x) \to \phi_1(x) \, \exp(-\imagi \int^t A_1(t')\,\diff t')$ here.

\bigskip

\subsection{Ground state}\label{sec:groundstate}
A time-dependent calculation most often starts from the ground state.  As we need anyway a code that propagates the RNO in time according \reff{tdphitilde} it would be convenient to use imaginary-time propagation for finding the ground state, as is commonly done in TDSE solvers.  On the other hand, it must be possible to derive the ground state in terms of RNO via a minimization approach. We show now that both ways will indeed lead to the same result. 

  The ground state energy to be minimized is
\begin{eqnarray}
\lefteqn{E = \int\diff 1 [\hat h_0(1') \gamma_1(1',1)]_{1'=1}} \nonumber \\
&& + \int\diff 1\!\!\int\diff 2\, \vee(|1-2|)\gamma_2(12,12).
\end{eqnarray}
Expressed in RNO, the energy becomes
\begin{eqnarray}
\lefteqn{E= 2   \sum_{i\ \mathrm{odd}} \bra{\tilde i} \hat h_0 \ket{\tilde i}_x}\nonumber \\
&&\quad  + \sum_{i\ \mathrm{odd}} \sum_{j\ \mathrm{odd}}  \frac{\eulere^{\imagi(\varphi_j-\varphi_i)}}{\sqrt{\braket{\tilde i}{\tilde i}_x \braket{\tilde j}{\tilde j}_x}} \, \bra{\tilde i\tilde i} \vee \ket{\tilde j\tilde j}_x ,\label{totenergyinrenormalizedNOs}
\end{eqnarray}
where we introduced the phases $\varphi_i$ via [cf.~Eq.\reff{Dimodsqu}]
\beq D_i=\sqrt{n_i} \,\eulere^{\imagi\varphi_i} \eeq
and explicitly write
\beq n_i =\braket{\tilde i}{\tilde i}_x. \eeq

We define a functional $\tilde E[\{\ket{\tilde i}\},\{\bra{\tilde i}\}]$ that takes 
the constraint $ \sum_{i\ \mathrm{odd}} \braket{\tilde i}{\tilde i} = \sum_{i\ \mathrm{odd}} n_i = 1$  via the Lagrange parameter $\epsilon$ into account, the orthogonality of the RNO via $\lambda_{ij}$,  $\lambda_{ii}=0$, the condition $n_i\geq 0$ via $\epsilon^0_i$, and $n_i\leq 1$ through $\epsilon^1_i$ (Karush-Kuhn-Tucker conditions, see, e.g., \cite{KKT,giesbthesis}), 
\begin{widetext}
\begin{eqnarray}
\tilde E &=&  2   \sum_{i\ \mathrm{odd}} \bra{\tilde i} \hat h_0 \ket{\tilde i} + \sum_{i\ \mathrm{odd}} \sum_{j\ \mathrm{odd}}  \frac{\eulere^{\imagi(\varphi_j-\varphi_i)}}{\sqrt{\braket{\tilde i}{\tilde i} \braket{\tilde j}{\tilde j}}} \, \bra{\tilde i\tilde i} \vee \ket{\tilde j\tilde j} \nonumber \\ && \qquad - \epsilon\left(\sum_{i\ \mathrm{odd}} \braket{\tilde i}{\tilde i}-1 \right)- \sum_{i\ \mathrm{odd}}\ \ \sum_{j\ \mathrm{odd}\ \neq \ i} \lambda_{ij} \braket{\tilde i}{\tilde j}    - \sum_{i\ \mathrm{odd}} [ \epsilon^0_i \braket{\tilde i}{\tilde i} + \epsilon^1_i (1- \braket{\tilde i}{\tilde i})]  .
\end{eqnarray}
Here, we dropped the index $x$ at $\langle \cdot | \cdot\rangle_x$.
The ``slackness conditions''  are  \cite{giesbthesis}
\beq \epsilon^0_i n_i = \epsilon^0_i \braket{\tilde i}{\tilde i} =0, \qquad \epsilon^1_i (1-n_i) = \epsilon^1_i(1- \braket{\tilde i}{\tilde i}) =0.\label{slackness} \eeq
Actually, the energy functional $\tilde E$ depends not only on the RNO but also on the phases  $\{\varphi_j\}$ of the geminal expansion coefficients. We suppress this dependence here because the values of these phases for the He spin-singlet groundstate case are  already known  [see Eq.~\reff{Dsforgroundstate} below]. A more general approach to the ground state problem based on geminals (where these phases are part of the minimization procedure) has been proposed and applied in \cite{mazziottiJChemPhys2000,MazziottiChemPhysLett2001}.    

Variation of $\tilde E$ with respect to $\bra{\tilde k}$ and $\ket{\tilde k}$ leads with 
\beq \epsilon_k=\epsilon+\epsilon^0_k-\epsilon^1_k\eeq
to 
\begin{eqnarray}\epsilon_k \ket{\tilde k} &=&\left\{ 2 \left[\hat h_0 +   \frac{1}{n_k}   \bra{\tilde k}   \vee \ket{\tilde k}(x)\right]   -  \frac{1}{n_k}  \Re\left[ \sum_{j\ \mathrm{odd}}    \frac{D_j D_k^* \bra{\tilde k \tilde k}  \vee \ket{\tilde j\tilde j}}{n_kn_j}\right]\right\} \ket{\tilde k} \nonumber \\
&& \qquad +  \sum_{j\ \mathrm{odd}\ \neq \ k} \left\{  2   \frac{D_jD_k^*}{n_j n_k} \, \bra{\tilde k}   \vee \ket{\tilde j}(x) -  \lambda_{kj} \right\} \ket{\tilde j}, \label{stationaryeq} 
\end{eqnarray}
and the hermitian conjugate of it. Multiplying by $\bra{\tilde i}$ from the left (and the hermitian conjugate by  $\ket{\tilde i}$ from the right) leads for $i=k$ to
\begin{eqnarray}
 \epsilon_i &=& \frac{1}{n_i} \left( 2  \bra{\tilde i}\hat h_0\ket{\tilde i}   +  \sum_{j\ \mathrm{odd}}      \left\{ 2\frac{D_jD_i^*}{n_in_j}\bra{\tilde i\tilde i}   \vee \ket{\tilde j\tilde j}  -  \Re \left[\frac{D_jD_i^*}{n_in_j}\bra{\tilde i\tilde i}   \vee \ket{\tilde j\tilde j} \right] \right\}\right), \label{epsiloni}\\
\forall i \qquad 0&=& \Im \sum_{j\ \mathrm{odd}}     \frac{D_jD_i^*}{n_j}\bra{\tilde i\tilde i}   \vee \ket{\tilde j\tilde j} , \label{conditiontobereal}
\end{eqnarray}
and for $i\neq k$ to
\begin{eqnarray}
\lambda_{ki} &=& \frac{2}{n_i} \left[ \bra{\tilde i}\hat h_0\ket{\tilde k}   +  \sum_{j\ \mathrm{odd}}     \frac{D_jD_k^*}{n_kn_j}  \, \bra{\tilde k\tilde i}   \vee \ket{\tilde j\tilde j}\right] = \lambda_{ik}^*,\label{lambdakiI}\\
\lambda_{ik} &=& \frac{2}{n_k} \left[\bra{\tilde k}\hat h_0\ket{\tilde i}   +  \sum_{j\ \mathrm{odd}}    \frac{D_j D_i^*}{n_in_j} \, \bra{\tilde i\tilde k}   \vee \ket{\tilde j\tilde j} \right].\label{lambdaikII}
\end{eqnarray}
From this follows
\begin{eqnarray}
 \bra{\tilde k}\hat h_0\ket{\tilde i} &=&   \frac{1}{n_i -n_k} \sum_{j\ \mathrm{odd}}   \frac{1}{n_j}\left(  D_k D_j^* \, \bra{\tilde j\tilde j}   \vee \ket{\tilde k\tilde i} -  D_j D_i^* \, \bra{\tilde i\tilde k}   \vee \ket{\tilde j\tilde j} \right),\label{khnulli}
\end{eqnarray}
which can be used to remove $\hat h_0$ entirely from the off-diagonal (with respect to the NO-index) part of the Hamiltonian. 

Putting Eqs.~\reff{stationaryeq}--\reff{khnulli} together we can write 
\beq 
0 = \hat\calH^{00}_n(x) \phitilde_n(x) + \sum_{k\ \mathrm{odd}\ \neq n}  \calH^{01}_{nk}(x) \phitilde_k(x), \label{tidepphitilde}\qquad \hat\calH^{00}_n(x) = A^0_n + \hat\calK^{00}_n(x), \quad \calH^{01}_{nk}(x) = B^0_{nk} + \calK^{01}_{nk}(x) 
\eeq
with
\begin{eqnarray}
A^0_n &=&  -  \frac{1}{n_n} \left( \bra{\tilde n}\hat h_0\ket{\tilde n} +  \Re\sum_{k\ \mathrm{odd}}     \frac{D_kD_n^*}{n_nn_k} \bra{\tilde n\tilde n}   \vee \ket{\tilde k\tilde k}\right) ,\quad \hat\calK^{00}_n(x) =\hat h_0 +     \frac{1}{n_n} \bra{\tilde n}   \vee \ket{\tilde n}(x), \label{an0}\\
B_{nk} &=& \frac{1}{n_k -n_n} \sum_{p\ \mathrm{odd}}   \left(    \frac{D_kD_p^*}{n_kn_p}\bra{\tilde p\tilde p}   \vee \ket{\tilde k\tilde n} -   \frac{D_pD_n^*}{n_nn_p} \, \bra{\tilde n\tilde k}   \vee \ket{\tilde p\tilde p} \right) , \quad \calK^{01}_{nk}(x)  =   \frac{D_kD_n^*}{n_kn_n} \, \bra{\tilde n}   \vee \ket{\tilde k}(x) .\label{K01nk}
\end{eqnarray}
\end{widetext}
Hence, we obtain by this variational method indeed the time-independent, ground state  analogue of Eqs.~\reff{tdphitilde}--\reff{k1nk}. This allows for finding the RNO ground state configuration (for a given set of OCN) by imaginary time-propagation of the time-dependent Schr\"odinger-like, nonlinear equation \reff{tdphitilde} (see Sec.~\ref{subsubsec:findingthegs} below).

The $\epsilon_k$ in \reff{stationaryeq} play the role of orbital energies. Unless the OCN are pinned to $n_k=0$ or $1$ we have non-integer $0<n_k<1$ for correlated systems. In such cases Eqs.~\reff{slackness} imply $\epsilon_k^0=0$, $\epsilon_k^1=0$ so that
\beq \epsilon_k=\epsilon,\label{allequal} \eeq
i.e., all orbital energies are equal. The ground state RNO and wavefunction expansion coefficients $D_j$ can be chosen real. Thus we have with \reff{epsiloni}
\beq \forall i \quad \epsilon=\epsilon_i=  \frac{1}{n_i} \left( 2  \bra{\tilde i}\hat h_0\ket{\tilde i}   +  \sum_{j\ \mathrm{odd}}   \frac{D_jD_i}{n_in_j}\bra{\tilde i\tilde i}   \vee \ket{\tilde j\tilde j}  \right)\eeq
and thus 
\beq \sum_{i\ \mathrm{odd}} \epsilon_i n_i = \epsilon \sum_{i\ \mathrm{odd}} n_i = \epsilon =E \label{everythingequal}\eeq
because the total energy is, according \reff{totenergyinrenormalizedNOs},
\beq E= 2   \sum_{i\ \mathrm{odd}} \bra{\tilde i} \hat h_0 \ket{\tilde i} + \sum_{i,j\ \mathrm{odd}} \frac{D_jD_i}{n_in_j} \, \bra{\tilde i\tilde i} \vee \ket{\tilde j\tilde j} .\label{totenerg}\eeq
As already noticed in \cite{gilbert}, the esthetically appealing result \reff{everythingequal} is puzzling, at least at first sight. All orbital energies are equal and equal the total energy of the system. Only with NO and their fractional OCN  the simple additive form  $E=\sum_i \epsilon_i n_i$ ---commonly known only from {\em non-interacting} systems---persists here for {\em interacting} systems.

\begin{figure}

\bigskip

\includegraphics[width=0.9\columnwidth]{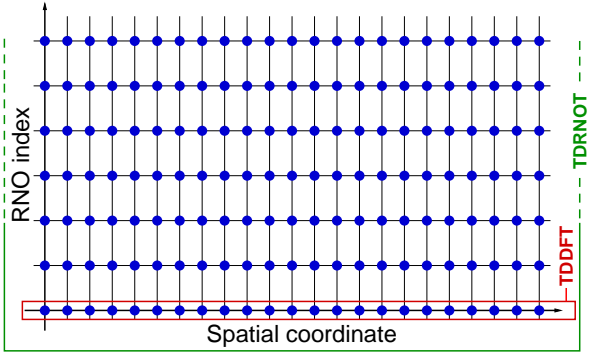} 
\caption{(color online). Sketch of the numerical grid. Each row corresponds to one of the RNO. Horizontal flux of probability density re\-pre\-sents motion in position space with the OCN kept constant. Flux in vertical direction implies a change in the OCN. Because we work with RNO instead of NO, the time-evolution on this grid is unitary. In the two-electron spin-singlet case the restriction to a single RNO (per spin) corresponds to TDHF or TDDFT in EXA approximation (red). In time-dependent RNO theory  (TDRNOT, green) more than one RNO are allowed.
 \label{fig:grid} }
\end{figure}

\subsection{Two orbitals per spin}
In our code that actually solves \reff{tdphitilde} we work on a numerical grid representing the discretized space variable $x$ and the RNO index $n=1,3, \ldots \NRNOmax$ (see Fig.~\ref{fig:grid}). All sums over the RNO indices have to be terminated at some finite $\NRNOmax$ in practice. Let us  consider the simplest yet nontrivial case of two RNO (per spin), i.e.,
\beq n_1=n_2\neq 0, \quad n_3=n_4\neq 0, \quad n_i=0 \ \mathrm{for} \  i>4, \eeq
which implies
\beq \tilde\phi_1=\tilde\phi_2, \quad  \tilde\phi_3=\tilde\phi_4, \quad \tilde\phi_i  \equiv 0 \ \mathrm{for} \ i>4. \eeq
As a single NO (per spin) is equivalent to TDHF (cf.\ Sec.~\ref{sec:tdhflimit}), allowing for two RNO per spin is just ``one small step beyond TDHF.'' However, we shall see below that two RNO per spin is already enough to describe some of the correlated two-electron dynamics completely missed by TDHF (or TDKS in EXA).
The relevant terms in \reff{tdphitilde} in this case are
\begin{eqnarray}
\hat\calH^0_1(x) &=& A_1 + \hat\calK^0_1(x),\quad\hat\calH^0_3(x) = A_3 + \hat\calK^0_3(x),\ \  \label{H0n2} \\
\calH^1_{13}(x) &=& B_{13} + \calK^1_{13}(x), \quad  \calH^1_{31}(x)={\calH^1_{13}}^*(x).
\end{eqnarray}
We find, making use of \reff{Dimodsqu},
\begin{widetext}
\begin{eqnarray}
A_1 &=&    - \frac{1}{n_1}\left[\Re\left(  \bra{\tilde 1\tilde 1}  \vee \ket{\tilde 1\tilde 1}  +  \frac{D_1\, D_3^*}{n_1n_3}  \bra{\tilde 3\tilde 3}  \vee \ket{\tilde 1\tilde 1}\right)
+ \bra{\tilde 1}\hat h_0 \ket{\tilde 1}\right] , \qquad A_3 = A_1[1\leftrightarrow 3],\label{A1}\\
\hat\calK^0_1(x) &=&  \hat h_0(x) +  \frac{\bra{\tilde 1} \vee \ket{\tilde 1}(x)}{n_1}, \qquad \hat\calK^0_3(x) = \hat\calK^0_1(x)[1\leftrightarrow 3] ,  \\
B_{13} &=&  \frac{1}{n_3-n_1}  \left( \frac{D_3\, D_1^*}{n_1n_3}
\bra{\tilde 1\tilde 1}  \vee \ket{\tilde 1\tilde 3}  + 
\frac{\bra{\tilde 3\tilde 3}  \vee \ket{\tilde 1\tilde 3}}{n_3} -  \left[
\frac{D_1\, D_3^*}{n_1n_3}
\bra{\tilde 3\tilde 3} \vee \ket{\tilde 3\tilde 1}  +  \frac{\bra{\tilde 1\tilde 1} \vee \ket{\tilde 3\tilde 1}   }{n_1}  \right]^*  \right), \quad B_{31} = B_{13}^*, \\
 \calK^1_{13}(x)  &=& \frac{D_3\, D_1^*}{n_1n_3}
\bra{\tilde 1} \vee \ket{\tilde 3}(x), \qquad \calK^1_{31}(x)  = {\calK^1_{13}}^*(x),\label{K13}
\end{eqnarray}
\end{widetext}
and the equation of motion has the simple structure
\beq \imagi\partial_t \twovec{\phitilde_1(x)}{\phitilde_3(x)} = \twotimestwo{ \hat\calH^0_1(x)}{ \calH^1_{13}(x)}{ \calH^1_{31}(x)}{ \hat\calH^0_3(x)}\twovec{\phitilde_1(x)}{\phitilde_3(x)}.  \label{eomtwoorbitalsmatrix}\eeq

Here, the off-diagonal elements determine whether the OCN are constant or not,
  \beq \dot n_1 = 2\Im  \bra{\tilde 1} \calH^1_{13}\ket{\tilde 3}, \qquad \dot n_3 = 2\Im  \bra{\tilde 3} \mbox{$\calH^1_{13}$}^*\ket{\tilde 1}. \label{n1dotn3dot} \eeq 
Furthermore, it is easy to show that the RNO $\tilde\phi_1$ and $\tilde\phi_3$ stay orthogonal at all times.

\section{Results and discussion} \label{sec:resanddisc}
The results in the following are obtained for the well-known model helium-like atom introduced in \cite{model1} and used extensively ever since (see, e.g., \cite{model2,ruggbau,appelgross,helbig}). In the spin-singlet configuration the full two-body TDSE reads 
\begin{eqnarray} \lefteqn{\imagi\pabl{}{t} \Psi(x_1x_2;t)}\label{fullTDSE}\\ 
&=& \left[ \sum_{i=1}^2\left( -\halb \pabl{^2}{x_i^2}  - \frac{2}{\sqrt{x_i^2 + 1}} - \imagi A(t)\pabl{}{x_i}\right)\right.  \nonumber \\
&& \quad\qquad   + \left. \frac{1}{\sqrt{(x_1-x_2)^2 + 1}}\right] \Psi(x_1x_2;t) .\nonumber \end{eqnarray}
 We employ velocity gauge to couple an external field with vector potential $A(t)$ in dipole approximation to the model atom (with the purely time-dependent term $\sim A^2$ ``transformed away'').
The corresponding one-body Hamiltonian for the RNO equations reads accordingly
\beq \hat h_0 =-\halb \pabl{^2}{x^2} - \imagi A(t)\pabl{}{x}  - \frac{2}{\sqrt{x^2 + 1}} \eeq
and
\beq \vee(x_1x_2)=\frac{1}{\sqrt{(x_1-x_2)^2 + 1}} . \eeq

For the real spin-singlet ground state wavefunction of the He atom the geminal expansion coefficients $D_i$ are
\beq D_1=\sqrt{n_1}, \qquad D_i=-\sqrt{n_i} \quad\mathrm{for}\quad i=3,5,7,\ldots\ , \label{Dsforgroundstate}\eeq
or, in terms of the phases introduced in Sec.~\ref{sec:groundstate},
\beq \varphi_1=0, \qquad \varphi_i=\pi \quad\mathrm{for}\quad i=3,5,7,\ldots\ . \eeq
A two-fermion system is the fortunate case where these phases are typically known for stationary configurations.
However, it is in general not known how the $D_i$ (or $\varphi_i$) change in time with an external driver $A(t)$ switched on (unless we solve the full two-electron TDSE and extract this information). 

\subsection{Ground state RNO} \label{subsubsec:findingthegs}
We applied imaginary-time propagation in combination with Gram-Schmidt orthogonalization to \reff{eomtwoorbitalsmatrix} with the $D_i$ according \reff{Dsforgroundstate} inserted in \reff{A1}--\reff{K13} in order to find the ground state configuration for given OCN. A grid with $N_x=500$ spatial points and a resolution of $\Delta x=0.4$ was found to be sufficient for that purpose. 

From Sec.~\ref{sec:groundstate} we know that if we try all combinations
\beq n_1=1.0-y, \quad n_3=y, \quad n_1+n_3=1, \quad y\in[0,1], \eeq
the ground state configuration will be the one for which
\beq \epsilon_1=\epsilon_3=\epsilon=E. \label{eq:crossingcondi}\eeq

\begin{figure}
\includegraphics[width=0.9\columnwidth]{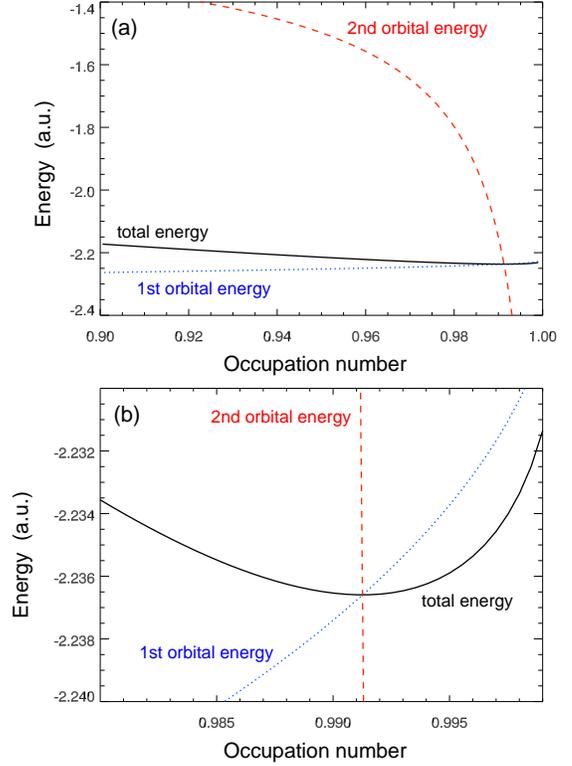} 
\caption{(color online). Orbital energies $\epsilon_1$ (dotted blue), $\epsilon_3$ (dashed red), and total energy $E$ (solid black) vs the dominant OCN $n_1$. The physically relevant RNO are obtained when the three curves cross [cf.\ \reff{allequal} and \reff{everythingequal}]. Panel (b) shows a zoom into the narrow $n_1$ region where all energies cross in one point. The curves  cross in the minimum of $E$.
 \label{fig1:energsvsOC0} }
\end{figure}

\begin{figure}
\includegraphics[width=0.9\columnwidth]{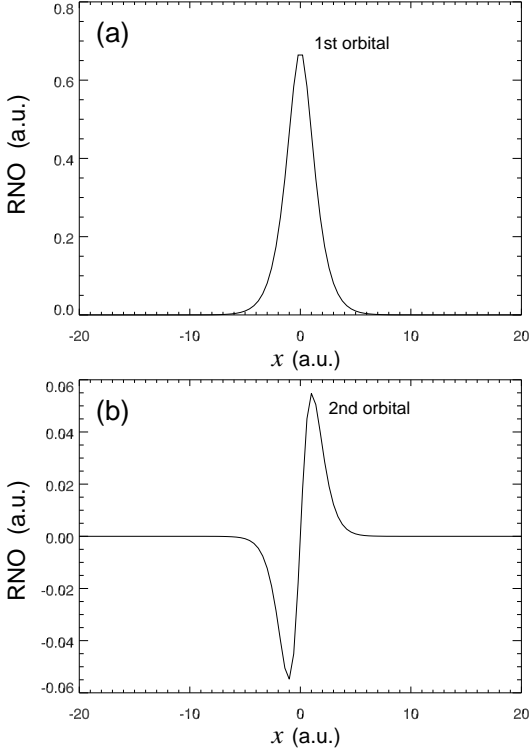} 
\caption{Ground state orbitals $\tilde \phi_1$ (a) and   $\tilde \phi_3$ (b) for $N_\mathrm{RNO}=2$. 
 \label{fig2:NOs} }
\end{figure}

Figure~\ref{fig1:energsvsOC0} shows  the orbital energies $\epsilon_1$, $\epsilon_3$ and the total energy $E$  according \reff{epsiloni} and \reff{totenerg}, respectively, vs $n_1$. Condition \reff{eq:crossingcondi} is fulfilled for
\beq n_1=0.99127, \qquad n_3=8.73\cdot 10^{-3},\eeq
with a total energy   \beq E_{N_\mathrm{RNO}=2}= -2.2366.  \eeq
The two NO are shown in Fig.~\ref{fig2:NOs}.

From the full TDSE solution on a spatial $500\times 500$-grid, also with $\Delta x=0.4$, we obtain the reference value for the ground-state energy
\beq   E_\mathrm{TDSE}= -2.2384  \eeq
and, via diagonalization of the 1RDM calculated from the two-body Schr\"odinger wavefunction,
for the first five OCN $n_1=0.99096$, $n_3=8.297\cdot 10^{-3}$, $n_5=7.063\cdot 10^{-4}$, $n_7=3.127\cdot 10^{-5}$, $n_9=7.392\cdot 10^{-6}$.
The relative error in the total energy is thus only 0.08\%. The HF ground state energy of the system is $E_\mathrm{HF}= 2.2243$, i.e., with $0.6$\% relative error. 

The OCN $n_1$ and $n_3$ we obtain for $N_\mathrm{RNO}=2$ are slightly above the exact values because the two alone are already forced to sum up to unity. 

The logarithmic ground state contour plot of the full TDSE two-body electron density $|\Psi(x_1x_2;t)|^2$ shows characteristic kinks along the diagonal $x_1=x_2$ (see Fig.~\ref{fig:TDSEspace}a below). These kinks are not reproduced with only two RNO since they are visible only  on a probability density level governed by higher-order RNO which have a  much smaller OCN and which are spatially more extended. Nevertheless, we will see that two RNO are already sufficient to describe, e.g.,  doubly-excited states and autoionization, not captured by (TD)DFT employing adiabatic xc-potentials.

\subsection{``Frozen'' Hamiltonian}
The full nonlinear system \reff{tdphitilde} [and even its truncated form \reff{eomtwoorbitalsmatrix}] is numerically difficult to handle because of instabilities. The time evolution of the RNO and the OCN (i.e., the RNO norms) is extremely sensitive to the phases $\varphi_i(t)$. Although we apply an unconditionally stable (i.e., exactly unitary) propagation algorithm (including predictor-corrector steps) to our Schr\"odinger-like equation \reff{eomtwoorbitalsmatrix} (with a hermitian Hamiltonian), the individual OCN $n_i(t)=\braket{\tilde i(t)}{\tilde i (t)}$---while still adding up to unity---tend to develop an unphysical, rather erratic behavior after some time of propagation. Such a behavior is actually to be anticipated from Eqs.~\reff{n1dotn3dot}: the phases of the $D_i$ {\em and} the phases of the RNO determine the changes in the OCN $n_i$ which, in turn, are fed back into the Hamiltonian. One way to mitigate these instabilities is to make approximations to $\tilde\gamma_{2,ijkl}$ that keep the OCN constant. However, the change of the OCN is often crucial to capture strongly driven, correlated, or resonant electron dynamics \cite{helbig,appelgross,rajam}.  

In TDKS calculations it is not uncommon to ``freeze'' the ground state Kohn-Sham potential during time-propagation, leading to the so-called ``bare'' Kohn-Sham response. The result can be compared to  the full TDKS calculation. In this way one may identify  the effect of the nonlinearity in the full Kohn-Sham potential. For instance, the peaks in the linear response spectrum obtained from the bare Kohn-Sham Hamiltonian correspond to the (allowed) transitions between the eigenstates of this Hamiltonian.  The  nonlinearity in the full Kohn-Sham potential typically moves the peaks from the frozen transition energies towards the correct position (see \cite{ruggbau} for the case of the model He atom studied here). We expect something similar for the differences in the linear response spectrum calculated with the full, nonlinear Hamiltonian in \reff{tdphitilde} [or, for two RNO, \reff{eomtwoorbitalsmatrix}] and the ground state RNO-frozen Hamiltonian. Let us denote the ground state RNO, OCN etc.\ by the corresponding underlined quantities. Then, for two RNO per spin, Eqs.~\reff{A1}--\reff{K13} become
\begin{widetext}
\begin{eqnarray}
\underline{A}_1 &=&    - \frac{1}{\underline{n}_1}\left[  \bra{\underline{\tilde 1\tilde 1}}  \vee \ket{\underline{\tilde 1\tilde 1}}  -  \frac{1}{\sqrt{n_1n_3}}  \bra{\underline{\tilde 3\tilde 3}}  \vee \ket{\underline{\tilde 1\tilde 1}}
+ \bra{\underline{\tilde 1}}\hat h_0 \ket{\underline{\tilde 1}}\right] ,\label{A1-} \quad \underline{\hat\calK}^0_1(x) =  \hat h_0(x) +  \frac{\bra{\underline{\tilde 1}} \vee \ket{\underline{\tilde 1}}(x)}{\underline{n}_1},  \\
\underline{B}_{13} &=&  \frac{1}{\underline{n}_3-\underline{n}_1}  \left(
\frac{\bra{\underline{\tilde 3\tilde 3}}  \vee \ket{\underline{\tilde 1\tilde 3}}}{\underline{n}_3} + \frac{\bra{\underline{\tilde 3\tilde 1}} \vee \ket{\underline{\tilde 3\tilde 3}}-\bra{\underline{\tilde 1\tilde 1}}  \vee \ket{\underline{\tilde 1\tilde 3}}}{\sqrt{\underline{n}_1\underline{n}_3}}
  - \frac{\bra{\underline{\tilde 3\tilde 1}} \vee \ket{\underline{\tilde 1\tilde 1}}   }{\underline{n}_1}   \right),  \quad \underline{\calK}^1_{13}(x)  = -\frac{1}{\sqrt{\underline{n}_1\underline{n}_3}}
\bra{\underline{\tilde 1}} \vee \ket{\underline{\tilde 3}}(x),\label{K13-}
\end{eqnarray}
\end{widetext}
where we have used  \reff{Dsforgroundstate}. The only remaining time-dependence in the Hamiltonian in \beq \imagi\partial_t \twovec{{\phitilde}_1}{{\phitilde}_3} = \twotimestwo{ \underline{\hat\calH}^0_1}{\underline{ \calH}^1_{13}}{\underline{ \calH}^1_{31}}{\underline{ \hat\calH}^0_3}\twovec{{\phitilde}_1}{{\phitilde}_3}  \label{eomtwoorbitalsmatrix-}\eeq 
then is in  $\hat h_0$ through the vector potential $A(t)$, i.e., only in the diagonal parts $\underline{\hat\calH}^0_1$ and $\underline{ \hat\calH}^0_3$.

Equations \reff{eomtwoorbitalsmatrix-}, \reff{A1-}--\reff{K13-} can be easily generalized to more than two RNO per spin. It is simple to prove that the sum of all OCN is conserved, $\partial_t \sum_i n_i=0$. With the frozen Hamiltonian the individual $n_i$ may {\em not} stay constant, which is a good feature, as remarked above. A drawback of the frozen Hamiltonian, however, is that $\partial_t \braket{\tilde i}{\tilde n} = 0$ is not strictly fulfilled anymore. This means that different RNO may not stay orthogonal during time propagation although, as eigenfunctions of a hermitian 1RDM, they should.

\subsection{Bare linear response}
\begin{figure}
\includegraphics[width=1.0\columnwidth]{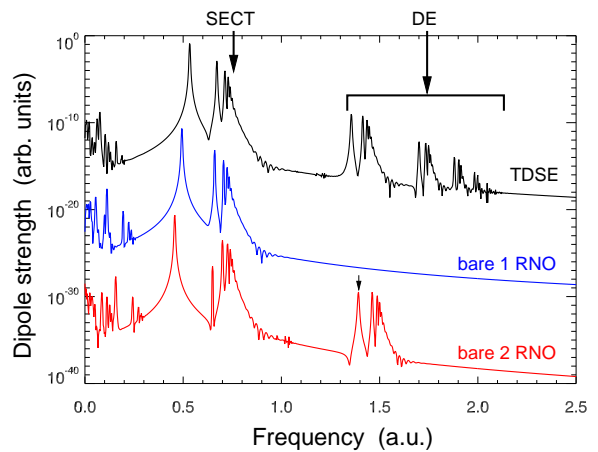} 
\caption{(color online). Linear response of the He model system obtained form the full TDSE \reff{fullTDSE} (upper, black, labeled ``TDSE''), with a single RNO per spin, i.e., equivalent to bare TDHF or TDKS in EXA (center, blue, labeled ``bare 1 RNO''), and two RNO (lower, red, labeled ``bare 2 RNO''). The single-electron continuum threshold is indicated by ``SECT'', doubly excited series of peaks by ``DE''. The little arrow indicates the first double-excitation peak in the ``bare 2 RNO''-result, discussed in Sec.~\ref{sec:autoioni}.
 \label{fig:linresp} }
\end{figure}

Figure~\ref{fig:linresp} shows linear response spectra for the He model system. All spectra were 
calculated by Fourier-transforming the dipole after disturbance of the system with a minute electric-field kick, corresponding to a step-like vector potential.  

The exact TDSE result shows the dominating series of peaks starting around a frequency $\omega=0.5$.  In the independent-electron picture, these peaks correspond to single-electron excitations: one electron stays in its ground state, the other one is excited to the first, second, etc.\ (dipole-allowed) state, up to the single-electron continuum threshold (SECT). Series of peaks corresponding to doubly-excited states follow around $\omega \simeq 1.3$ and greater. As in the ``real'' He atom, all double-excitations are embedded in the single-ionization continuum and thus are autoionizing. Ultimately, with excitations of frequencies $\omega>E$, both electrons can be lifted into the two-electron continuum \cite{remarkonlowfrequpeaks}.

The linear response spectrum obtained  with just one RNO (per spin) is equivalent to the result of a bare TDHF (or TDKS-EXA) calculation \cite{remarkonfullTDKSspec}. One can show that doubly-excited states are not even covered by full TDDFT (i.e., without frozen Kohn-Sham potential)  as long as adiabatic exchange-correlation (xc) potentials are employed (see, e.g., \cite{ullrich}). With a frozen Hamiltonian and just a single RNO it is immediately clear that double-excitations cannot exist. And indeed, the peaks corresponding to transitions to doubly-excited states are absent for the ``bare 1 RNO''-result in Fig.~\ref{fig:linresp}. The single-excitation series is there but a bit red-shifted as compared to the TDSE reference spectrum.

The first main result of this paper is that the first series of double-excitation peaks is present when Eq.~\reff{eomtwoorbitalsmatrix-} is solved. In order to reproduce double-excitations in TDDFT one would have to use xc potentials with memory \cite{maitraautoioni}. Even if useful potentials with memory were known they would very likely be computationally expensive. Instead, with our time-dependent RNO theory (TDRNOT) we cover double-excitations even with a frozen Hamiltonian and just two RNO. This is because we allow for more than one orbital per spin (or, in the single-particle picture, per particle), like in a configuration interaction calculation or in multi-configurational Hartree-Fock. However, the advantage of TDRNOT over such methods is that the RNO constitute automatically the most adequate time-dependent basis, as mentioned already in the Introduction. This is the reason why we get along with only two RNO. Thanks to this small number of necessary RNO we have only  little computational overhead compared to the  corresponding TDHF or TDKS-EXA simulation. 

The single-excitation peaks in the ``bare 2 RNO'' spectrum are even more red-shifted compared to the TDSE than in the ``bare 1 RNO'' result, while the peaks corresponding to double-excited states are slightly blue-shifted. We have checked that with three RNO the next series of double excitation peaks (seen in the TDSE result starting around $\omega\simeq 1.7$) is also reproduced, although even more blue-shifted.  ``Unfreezing'' of the Hamiltonian should cure these shifts and improve the quantitative agreement between TDSE and TDRNOT results but, as mentioned above, we first have to overcome the instability problems before we can check this assertion.

\subsection{Autoionization after excitation by a laser pulse}\label{sec:autoioni}
\begin{figure}
\includegraphics[width=0.9\columnwidth]{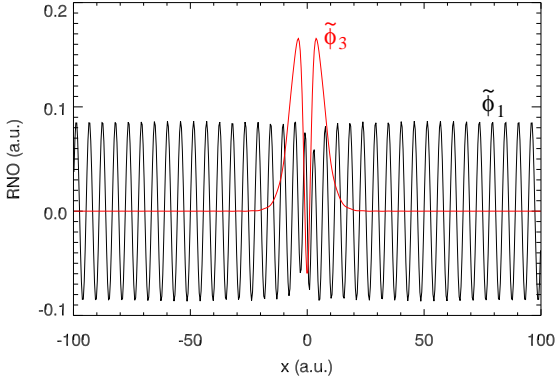} 
\caption{(color online). The two RNO $\phitilde_1(x)$ and $\phitilde_3(x)$ of the autoionizing state  indicated by the small vertical arrow in  Fig.~\ref{fig:linresp} for the bare 2 RNO calculation. 
 \label{fig:autoioniRNO} }
\end{figure}

The small arrow in Fig.~\ref{fig:linresp} indicates the lowest-lying transition to a doubly-excited state in the bare TDRNOT He-system employing two RNO. The two RNO for that particular state were calculated via diagonalization of the bare Hamiltonian in \reff{eomtwoorbitalsmatrix-} on a grid with $500$ spatial grid points. The result is depicted in Fig.~\ref{fig:autoioniRNO}. The dominating RNO $\phitilde_1(x)$ has an OCN $n_1=0.733$ and is delocalized, thus allowing for outgoing electron flux. Indeed, the spatial oscillations fit to the wavenumber $k=\sqrt{2(\omega- \Ip)}\simeq 1$  expected for the outgoing electron (with $\omega\simeq 1.38$ the energy required to populate the autoionizing state and the ionization potential inferred from the SECT in  Fig.~\ref{fig:linresp}, $\Ip\simeq 0.75$). The second RNO $\phitilde_3(x)$ is localized close to the origin and has an OCN $n_3=0.267$. If one detunes from the autoionizing resonance the OCN of the delocalized NO increases. Hence, the characterizing feature of an autoionizing resonance (as compared to states in a ``flat'' continuum) is the relative increase of the occupation of localized NO. 

As an example for a TDRNOT calculation beyond linear response we have simulated the interaction of the model He atom with an $\Ncyc=30$-cycle  $\sin^2$-shaped laser pulse of frequency $\omega$, i.e.,
\beq A(t) = \Ahat\, \sin^2\left(\frac{\omega t}{2 \Ncyc} \right)\, \sin \omega t \label{laserpulse}\eeq
for $0<t<\Ncyc T$ with $T=2\pi/\omega$, and zero otherwise. 
The frequency in one calculation was chosen resonant with the transition to the first doubly-excited, autoionizing state while in another run, for comparison, we tuned it off-resonant. In all cases the vector potential amplitude was $\Ahat=0.01$.

\subsubsection{TDSE result in position space}
Before showing the results obtained with the bare TDRNOT equation \reff{eomtwoorbitalsmatrix-}, let us illustrate the exact dynamics we observe by solving the full two-electron TDSE \reff{fullTDSE}. The same model atom has been employed to study autoionization in the presence of an additional, optical laser pulse in \cite{zhao}. 

\begin{figure}
\includegraphics[width=1.0\columnwidth]{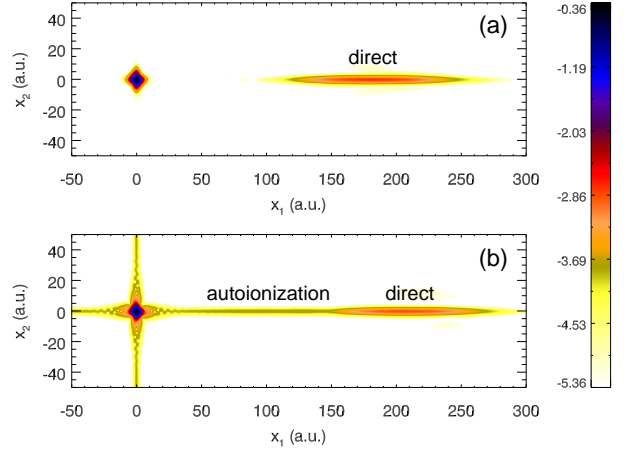} 
\caption{(color online). Logarithmic probability density (over five orders of magnitude) from \reff{fullTDSE} for a $\Ncyc=30$-cycle laser pulse with $\Ahat=0.01$ at $t=255$ for the off-resonant case $\omega=1.25$ (a) and the resonant one  $\omega=1.36$ (b). In (b), ionization continues after the laser pulse, as can be seen from the probability density trailing the (directly) laser-generated photoelectron wavepacket.
 \label{fig:TDSEspace} }
\end{figure}

For the off-resonant case we chose $\omega=1.25$. In this case, we expect single-photon ionization while the laser is on and no ionization thereafter. In fact, this can be clearly inferred from the (logarithmically scaled) position space probability density plotted in Fig.~\ref{fig:TDSEspace}a  for time $t=255$, i.e., well after the laser was off at $30T=151$. Around the origin, the remaining ground state density is visible. Around $x_1\simeq 180$, $x_2=0$ the laser-generated photoelectron wavepacket is seen. It travels to the right with velocity (or wavenumber) $k=\sqrt{2(\omega- \Ip)}\simeq 1$, where the ionization potential (for the TDSE calculation) is $\Ip=0.751$. Analogous wavepackets travel in the $-x_1$ and $\pm x_2$ directions (not shown).

The frequency $\omega=1.36$ for the resonant case can be read off Fig.~\ref{fig:linresp}. Figure~\ref{fig:TDSEspace}b shows the probability density, again at time  $t=255$. Because the frequency is higher than in (a), the photoelectron wavepacket is faster and narrower ($\Ncyc$ was kept constant). The major qualitative differences compared to the off-resonant case is that (i) the system continues to ionize after the laser is off, as is seen from the trailing edge of probability density following the (directly) laser-generated wavepacket, and (ii) the different structure in the probability density around the origin. The latter clearly shows that the system is not left in the ground state after the interaction with the laser pulse. Instead, it is in a superposition of ground and autoionizing state. Similar patterns are shown in \cite{zhao}.

\subsubsection{Bare two-RNO result in position space}
\begin{figure}
\includegraphics[width=1.0\columnwidth]{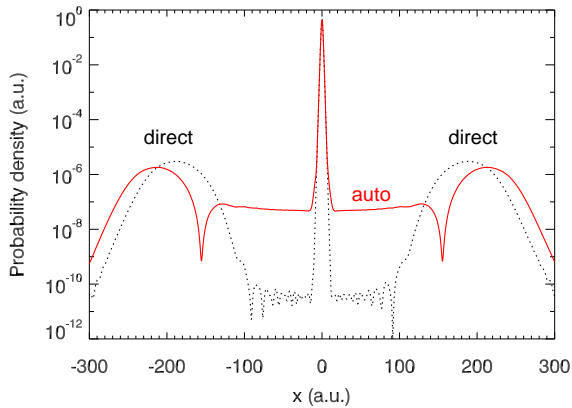} 
\caption{(color online). TDRNOT result corresponding to the TDSE two-electron density dynamics in Fig.~\ref{fig:TDSEspace}. The off-resonant frequency used was $\omega=1.28$ (other parameters as in the TDSE runs), leading to the black, dotted density at $t=255$. The resonant frequency $\omega=1.38$ yields the red, solid  density.
 \label{fig:TDRNOTspace} }
\end{figure}

The probability density $ |\tilde\phi_1(x)|^2 + |\tilde\phi_3(x)|^2$ from the bare  two-RNO simulation at $t=255$ is shown in Fig.~\ref{fig:TDRNOTspace}. The laser frequency for the resonant excitation of the autoionizing state was tuned to the respective value $\omega=1.38$ indicated by an arrow in Fig.~\ref{fig:linresp}.  For the non-resonant run $\omega=1.28$ was chosen. Again, left and right-going photoelectron wavepackets (with their maxima at the expected positions) are observed. In  the case with autoionization (red, solid) a much higher probability density level between the origin and the wavepackets is observed. This is the 1D analogue of the TDSE density dynamics  in the $x_1,x_2$-plane in Fig.~\ref{fig:TDSEspace}.

\subsubsection{Momentum spectra}
With autoionization involved, the electron spectra should display Fano line shapes \cite{fano}. Strictly speaking, one should calculate photoelectron spectra by projecting the final wavefunction at a time when the laser is off on field-free continuum eigenstates. When the final wavefunction is not available but only the RNO one can, in principle, rewrite the expectation value of the corresponding spectral projection operator in terms of the 2DM $\gamma_2$ or the coefficients $D_i$. The latter need to be approximated anyway in the general TDRNOT approach. Hence, there is no conceptual problem in calculating any observable for systems with two-body interactions because the knowledge of $\gamma_2$ suffices. Note that in TDDFT it is not clear how to calculate photoelectron spectra in a rigorous way  because the density functional for this observable is unknown \cite{TDDFTspec}.  

\begin{figure}
\includegraphics[width=0.9\columnwidth]{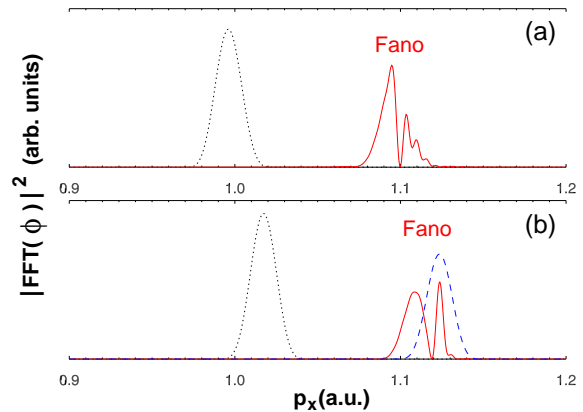} 
\caption{(color online). Linearly scaled photoelectron momentum spectra obtained by Fourier-transforming (FFT) the outgoing part of the TDSE wavefunction (a) and RNO (b) for a $\Ncyc=100$ laser pulse of the form \reff{laserpulse} with $\Ahat=0.01$. The same resonant (solid, red) and off-resonant (dotted, black) frequencies as in Figs.~\ref{fig:TDSEspace} and \ref{fig:TDRNOTspace} were chosen. The spectra were obtained at times $t=750$.  Fano profiles are seen in the TDSE result at resonance and the  TDRNOT result at resonance when 2 RNO are employed (red).  In panel (b) the result for a single NO and $\omega=1.38$ (dashed, blue) is also plotted, showing no Fano profile.
 \label{fig:momspecs} }
\end{figure}

Projection on field-free continuum states is computationally expensive. Hence, it is quite common in practice to project-out the most populated bound states and Fourier-transform the ``rest'' in order to obtain momentum spectra. Alternatively, one may filter-out the region around the origin. While not being a rigorous way to calculate photoelectron spectra, the method yields sufficiently accurate spectra for our purposes.
Hence, we pursued the same strategy and Fourier-transformed the outgoing part of the wavefunction. In this way we  obtain the momentum spectra shown in Fig.~\ref{fig:momspecs}. 

The TDSE results for the modulus-square of the Fourier transform (integrated over $p_{x_2}$) are presented in  Fig.~\ref{fig:momspecs}a. The off-resonant excitation by an $\Ncyc=100$ laser pulse of the form \reff{laserpulse} (same frequency $\omega=1.25$ and field amplitude as in Fig.~\ref{fig:TDSEspace}) leads to a peak close to momentum $p_x=1$, as expected. The resonant excitation leads to a Fano ``kink'' in the photoelectron peak at $p_x\simeq 1.1$ 

Figure~\ref{fig:momspecs}b shows the bare 2 RNO TDRNOT-results. The second main result of this work---after the existence of doubly-excited states---is that our approach  also yields a resonance peak with a Fano kink (solid, red). Although the detailed shape of the Fano resonance differs from the TDSE result it is remarkable that it is present at all. As emphasized already, autoionizing states are not captured by a TDHF or TDKS calculation (with adiabatic xc potentials). As a consequence,  Fano profiles will not be present there either. The black, dotted peak in   Fig.~\ref{fig:momspecs}b is the result for the off-resonant frequency. The dashed blue peak is obtained for $\omega=1.38$ but with one NO only (equivalent to the bare TDHF result). As there are no doubly excited states with one NO, there is no autoionization and thus no Fano profile.

\section{Conclusion and outlook} \label{sec:concl}
We have introduced time-dependent renormalized natural orbital theory (TDRNOT) and tested it with a numerically exactly solvable model helium atom. The main result is that even with only two renormalized natural orbitals (RNO) and the bare (i.e., ``frozen'') effective ground state Hamiltonian we observe correlation signatures impossible to capture with time-dependent density functional theory (TDDFT) using adiabatic exchange-correlation potentials, namely  (i) excitation of doubly excited states in the linear response spectrum and---beyond linear response---(ii) autoionization and Fano profiles in the photoelectron spectra. The numerical effort scales with the number of RNO cubed but only linearly with the number of spatial gridpoints
required for one particle.
While in TDDFT exchange-correlation potentials with memory are required to capture  doubly excited states, the effective Hamiltonian in TDRNOT is local in time. Moreover, the problem in TDDFT concerning density functionals for observables that are not explicitly known in terms of the single-particle density is removed.

Future work will be devoted to other two-electron systems in laser fields (in full dimensionality, spin-triplet, H$_2$), resonant interactions, and three-electron model systems that are still numerically exactly solvable. In full dimensionality, each RNO will be expanded in spherical harmonics and with the radial coordinate discretized as in this work. Then, one may apply analogs of the central field approximation for ground state calculations and a multipole expansion of the effective Hamiltonian for time-dependent simulations. In the spin-triplet case the algebraic structure of the two-body density matrix (or two-electron wavefunction) expansion is different from the spin-singlet but otherwise the approach is the same. The same holds for H$_2$. Three-electron systems such as the lithium atom in intense laser fields is of interest as it is the simplest system with ``inner'' electrons in a closed shell. The time-dependent Schr\"odinger equation of 1D model Li atoms in intense laser fields is still possible to solve exactly \cite{Li} and thus may serve as a benchmark for our method. Another research direction will be the development of density matrix functionals that allow to go beyond the bare TDRNOT without unleashing instabilities in the RNO equation of motion.

\section*{Acknowledgment}
This work was supported by the SFB 652 of the German Science Foundation (DFG).

\begin{appendix}

\begin{widetext}
 
\section{Derivation of EOM \reff{eomrenormphis}} \label{sec:derivationI}
Inserting \reff{gamma1rdm} and \reff{gamma2rdm} into \reff{eom1rdm} yields (suppressing all time arguments)
\begin{eqnarray}
\lefteqn{-\imagi \sum_k [\partial_t n_k] \phi_k^*(1') \phi_k(1) -\imagi \sum_k n_k [\partial_t\phi_k^*(1')] \phi_k(1) -\imagi \sum_k n_k \phi_k^*(1')[\partial_t \phi_k(1)]}\label{tmpi} \\
&=&  [\hat h_0(1') -  \hat h_0(1)] \sum_k n_k \phi_k^*(1') \phi_k(1)  + 2 \int \diff 2 \,  \bigl\{ \vee(|1'-2|) - \vee(|1-2|) \bigr\}  \sum_{ijkl} \gamma_{2,ijkl} \phi_i(1) \phi_j(2) \phi_k^*(1') \phi_l^*(2) .\nonumber
\end{eqnarray}
Because the NO form a complete basis we may introduce time-dependent coefficients $\alpha_{km}(t)$ \cite{appelthesis} such that
\beq \imagi \partial_t \phi_k(t) = \alpha_{kk}(t) \phi_k(t) + \sum_{m\neq k} \alpha_{km}(t) \phi_m(t) . \label{alphas} \eeq
Here we drop the argument $1$ as the equation holds for all position and spin degrees of freedom. 
With the transformation
\beq \phi_k(t) = \eulere^{ -\imagi \int^t \alpha_{kk}(t')\,\diff t' } \phi'_k(t)  \eeq
[where $\alpha_{kk}$ is real because of $\alpha_{km}(t) = \alpha_{mk}^*(t)$]
we transform-away the diagonal part of $\alpha_{km}$  and obtain
\beq  \imagi \partial_t  \phi'_k(t)   = \sum_{m\neq k}\alpha'_{km}(t) \phi'_m(t) \label{eq:phitilde} , \qquad \alpha'_{km}(t) = \alpha_{km}(t)  \eulere^{ -\imagi \int^t [\alpha_{mm}(t')-\alpha_{kk}(t')]\,\diff t' }. \eeq
$\alpha'_{km}$ constitutes another hermitian matrix. The NO are defined as the eigenfunctions of the hermitian 1RDM. 
We have the freedom to choose the phase in such a way that \reff{eq:phitilde} holds. This is the analogue of switching to the interaction picture, but now for a nonlinear Hamiltonian. Dropping the primes, $ \phi' \to \phi$, $\alpha' \to \alpha$, 
we have for \reff{tmpi}
\begin{eqnarray}
 \lefteqn{ -\imagi \sum_k [\partial_t n_k] \phi_k^*(1') \phi_k(1) +  \sum_{k,m\neq k} n_k \alpha_{km}^* \phi_m^*(1') \phi_k(1) - \sum_{k,m\neq k} n_k \phi_k^*(1')\alpha_{km} \phi_m(1) } \\
 &=&  [\hat h_0(1') -  \hat h_0(1)] \sum_k n_k \phi_k^*(1') \phi_k(1) + 2 \int \diff 2 \,  \bigl\{ \vee(|1'-2|) - \vee(|1-2|) \bigr\}  \sum_{ijkl} \gamma_{2,ijkl} \phi_i(1) \phi_j(2) \phi_k^*(1') \phi_l^*(2) . \nonumber
\end{eqnarray}
Multiplying by $\phi_n(1')$, $\phi_p^*(1)$, and integrating over $1$ and $1'$ gives
\beq -\imagi  \dot{n}_n \delta_{np}  + n_p \alpha_{pn}^*  -  n_n \alpha_{np}  = (n_p-n_n)\bra{p} \hat h_0 \ket{n}  + 2 \sum_{jkl} \left\{ \gamma_{2,pjkl} \bra{kl}  \vee \ket{nj}   -  \bigl[\gamma_{2,njkl}  \bra{kl} \vee \ket{pj}   \bigr]^*  \right\} . \label{eomndot}\eeq
For $n=p$ we obtain  \reff{ndotn},
\beq  \dot{n}_n    =   - 4  \Im  \sum_{jkl} \gamma_{2,njkl} \bra{kl}  \vee \ket{nj}   . \label{eq:ndotappendix}\eeq
For $n\neq p $ we have
\beq \alpha_{np} ( n_p - n_n ) = (n_p-n_n)\bra{p} \hat h_0 \ket{n}  + 2 \sum_{jkl} \left\{ \gamma_{2,pjkl} \bra{kl}  \vee \ket{nj}   -  \bigl[\gamma_{2,njkl}  \bra{kl} \vee \ket{pj}   \bigr]^*  \right\} .  \eeq
If $n_p\neq n_n$ we can divide by $n_p-n_n$, which gives with \reff{eq:phitilde}
\beq \imagi \partial_t \phi_n =   \sum_{p\neq n}  \left( \bra{p} \hat h_0 \ket{n}  + \frac{2}{n_p-n_n} \sum_{jkl} \left\{ \gamma_{2,pjkl} \bra{kl}  \vee \ket{nj}   -  \bigl[\gamma_{2,njkl}  \bra{kl} \vee \ket{pj}   \bigr]^*  \right\}\right) \phi_p. \label{NOeom} \eeq
Equations \reff{NOeom} and \reff{eomndot} are EOM for NO and OCN, respectively, but not useful for our purposes because truncating them to a small number of NO would yield extremely  poor results \cite{poorresultremark}. 
Instead, it is more suitable to seek for an EOM of the form of Eq.~\reff{eq:desiredform},
$ \imagi\partial_t \boldsymbol\Phi(1) = \hat{{\cal H}}(1) \boldsymbol\Phi(1), $
where $\boldsymbol\Phi(1,t)$ is a vector of NO $\phi_k(1,t)$ and $\hat{{\cal H}}$ acts in NO and position-spin space (see Fig.~\ref{fig:grid}) instead of only in NO space as in  \reff{NOeom}. Multiplying \reff{tmpi} by $\phi_n(1')$ and integrating-out only $1'$ (but not $1$) leads to
\begin{eqnarray}
\imagi \partial_t \phi_n(1) &=& \frac{4\imagi}{n_n}   \left\{ \Im  \sum_{jkl} \gamma_{2,njkl} \bra{kl}  \vee \ket{nj}\right\}\phi_n(1)-  \bra{n}\hat h_0 \ket{n} \phi_n(1) \nonumber\\
&& + \sum_{k\neq n}  \frac{2n_k/n_n}{n_k-n_n} \sum_{jpl} \left\{ \gamma_{2,kjpl} \bra{pl}  \vee \ket{nj}   -  \bigl[\gamma_{2,njpl}  \bra{pl} \vee \ket{kj}   \bigr]^*  \right\}\phi_k(1) - \frac{2}{n_n}\sum_{kjpl} \gamma_{2,kjpl}  \bra{pl} \vee \ket{nj} \, \phi_k(1)\nonumber \\
&&   + \hat h_0(1) \phi_n(1) + \frac{2}{n_n} \sum_{kjl} \gamma_{2,kjnl} \bra{l} \vee \ket{j}(1) \, \phi_k(1) .\label{eomspaceandNO}
\end{eqnarray}
In the first line on the right hand side there are purely time-dependent coefficients in front of $\phi_n(1)$. This part of the total Hamiltonian $ \hat{{\cal H}}(1)$ is diagonal in both NO-index and position-spin space.
In the second line we have the terms corresponding to the NO-index off-diagonal part of the Hamiltonian.
In the third line the first term does the ``usual'' coupling in position (-spin) space due to the single-particle Hamiltonian $\hat h_0$. This term is diagonal in NO-space. The second term in the third line  couples the NO  through space-dependent (though multiplicative) coefficients.

It is simple to rewrite the EOM  \reff{eomspaceandNO} in terms of the RNO \reff{phitildes}.
Using  \reff{eq:ndotappendix} and \reff{eq:gammatilde},
the result \reff{eomrenormphis} is obtained.

\section{Connection between $\gamma_{2,njkl}$ and $D_i$} \label{app:gamma2andD}
For the 2DM in the two-electron spin-singlet case we find, using \reff{twoelspinsingletwf}, 
\beq \gamma_2(12,1'2') = \Phi(x_1x_2) \Phi^*(x_1'x_2')\, \frac{1}{2} (\delta_{\sigma_1+}\delta_{\sigma_2-} - \delta_{\sigma_1-}\delta_{\sigma_2+} ) (\delta_{\sigma_1'+}\delta_{\sigma_2'-} - \delta_{\sigma_1'-}\delta_{\sigma_2'+} ) , \eeq
and with \reff{gamma2rdm}
\begin{eqnarray} \lefteqn{\Phi(x_1x_2) \Phi^*(x_1'x_2')\, \frac{1}{2} (\delta_{\sigma_1+}\delta_{\sigma_2-} - \delta_{\sigma_1-}\delta_{\sigma_2+} ) (\delta_{\sigma_1'+}\delta_{\sigma_2'-} - \delta_{\sigma_1'-}\delta_{\sigma_2'+} )} \\
&& \qquad\qquad\qquad =  \sum_{ijkl} \gamma_{2,ijkl} \phi_i(x_1)\delta_{\sigma_1\sigma_i} \phi_j(x_2)\delta_{\sigma_2\sigma_j} \phi_k^*(x_1')\delta_{\sigma_1'\sigma_k} \phi_l^*(x_2')\delta_{\sigma_2'\sigma_l}. \nonumber
\end{eqnarray}
Multiplication by NO and integration allows to solve for the $\gamma_{2}$ expansion coefficients,
\begin{eqnarray} \gamma_{2,mnop}
 &=& \frac{1}{2} (\delta_{\sigma_o+}\delta_{\sigma_p-} - \delta_{\sigma_o-}\delta_{\sigma_p+} )\left( \delta_{\sigma_m +}\delta_{\sigma_n-} -\delta_{\sigma_m -}\delta_{\sigma_n+} \right) \ \gamma^{(x)}_{2,mnop} \label{gammamnopsinglet} \end{eqnarray}
with the spatial part
\begin{eqnarray} \gamma^{(x)}_{2,mnop}&=&\int\diff x_1\int\diff x_2\int\diff x_1'\int\diff x_2'\,  \phi^*_n(x_2) \phi^*_m(x_1) \Phi(x_1x_2) \Phi^*(x_1'x_2')\phi_o(x_1') \phi_p(x_2')  \\
&=& \int\diff x_1\int\diff x_2\int\diff x_1'\int\diff x_2'\,  \phi^*_n(x_2) \phi^*_m(x_1)  \sum_{i \ \mathrm{odd}}  D_i\, \phi_i(x_1)\phi_i(x_2) \sum_{j \ \mathrm{odd}}  D_j^*\, \phi_j(x_1')\phi_j(x_2')\phi_o(x_1') \phi_p(x_2').\nonumber \end{eqnarray}
Now, one must not forget  that, e.g.,  not only $m=i$ contributes but also $m=i+1$ because $\phi_i=\phi_{i+1}$ for $i$ odd. Hence,
\beq
\gamma^{(x)}_{2,mnop}
= (\delta_{m,i} + \delta_{m,i+1})(\delta_{n,i} + \delta_{n,i+1}) D_i\, D_j^*\, (\delta_{o,j} + \delta_{o,j+1})(\delta_{p,j} + \delta_{p,j+1}) \quad\mathrm{for} \ i,j \ \mathrm{odd} \label{gamma2xmnop}
\eeq
and thus
\beq \gamma_{2,mnop} = \frac{1}{2} (\delta_{\sigma_o+}\delta_{\sigma_p-} - \delta_{\sigma_o-}\delta_{\sigma_p+} )\left( \delta_{\sigma_m +}\delta_{\sigma_n-} -\delta_{\sigma_m -}\delta_{\sigma_n+} \right)\qquad\qquad\qquad\qquad\qquad\qquad\qquad\qquad\qquad  \label{gamma2mnop}\eeq
\[\qquad \qquad \qquad \times  (\delta_{m,i} + \delta_{m,i+1})(\delta_{n,i} + \delta_{n,i+1}) D_i\, D_j^*\, (\delta_{o,j} + \delta_{o,j+1})(\delta_{p,j} + \delta_{p,j+1}) \quad\mathrm{for} \ i,j \ \mathrm{odd}  . \]

Because of the spin-part and our indexing (odd index $\leftrightarrow$ spin-up, even index  $\leftrightarrow$   spin-down, see Sec.~\ref{sec:twoelspsing}), for $\gamma_{2,mnop}$ not to vanish the index pair  $(o,p)$ must be (even, odd) or (odd, even). The same holds for the index pair ($m,n$).

In the case of two NO per spin this leads to the nonvanishing $\gamma_{2,mnop}$ summarized in Table~\ref{table:gamma2}. We not only have the general property $\gamma_{2,mnop}=\gamma_{2,opmn}^*$ here but also $ \gamma_{2,mnop}= - \gamma_{2,mnpo} = - \gamma_{2,nmop} = \gamma_{2,nmpo}$. We also show the results for typical approximations of the form \reff{gamma2approx}, which give erroneously zeros for ``cross-block'' elements like $mnop=1234$ or $3421$ while they may give erroneously diagonal contributions, e.g.,  $f_{1111}-g_{1111}$ unless $f_{mnop}=g_{mnop}$, as in Hartree-Fock. 

\end{widetext}

\begin{table}
\caption{Nonvanishing  $\gamma_{2,mnop}$ in the case of two NO per spin in the two-electron spin-singlet case and for approximations of the form  \reff{gamma2approx}. \label{table:gamma2}}
\begin{tabular}{cccccc}\hline\hline
$m$  & $n$ & $o$ & $p$ & $\gamma_{2,mnop}$ &  $\gamma^\mathrm{(approx)}_{2,mnop}\qquad$  \\ \hline
1   &  2  & 1  &  2  &    $\frac{1}{2}  |D_1|^2 = n_1/2$     &  $f_{1212}$  \\
1   &  2  &  2  &  1  &    $-\frac{1}{2} |D_1|^2 = -n_1/2$     & $-g_{1221}$   \\
1   &  2  &  3 &  4  &    $\frac{1}{2}  D_1D_3^* = \halb \sqrt{n_1 n_3}\, \eulere^{\imagi\varphi}$     & 0    \\
1   &  2  &  4 &   3 &    $-\frac{1}{2}  D_1D_3^* = -\halb \sqrt{n_1 n_3}\, \eulere^{\imagi\varphi}$     & 0  \\ \hline

2   &  1  & 1 & 2 &  $-\frac{1}{2} |D_1|^2= -n_1/2$      & $-g_{2112}$   \\
2   &  1  & 2 & 1 &  $\frac{1}{2} |D_1|^2 = n_1/2 $     &   $f_{2121}$ \\
2   &  1  & 3 & 4 &  $-\frac{1}{2} D_1 D_3^* = -\halb \sqrt{n_1 n_3}\, \eulere^{\imagi\varphi}$     & 0  \\
2   &  1  & 4 & 3 &  $\frac{1}{2}D_1 D_3^*= \halb \sqrt{n_1 n_3}\, \eulere^{\imagi\varphi}$     & 0  \\ \hline

3   &  4  & 1 & 2& $\frac{1}{2} D_3 D_1^*= \halb \sqrt{n_1 n_3}\, \eulere^{-\imagi\varphi}$     &  0  \\
3   &  4  & 2 & 1& $-\frac{1}{2} D_3 D_1^*= -\halb \sqrt{n_1 n_3}\, \eulere^{-\imagi\varphi}$     & 0  \\
3   &  4  & 3 & 4& $\frac{1}{2} |D_3|^2 = n_3/2$     & $f_{3434}$   \\
3   &  4  & 4 & 3& $-\frac{1}{2} |D_3|^2=-n_3/2$     & $-g_{3443}$  \\ \hline

4   &  3  & 1 & 2 &   $-\frac{1}{2}  D_3 D_1^*= -\halb \sqrt{n_1 n_3}\, \eulere^{-\imagi\varphi}$   & 0  \\ 
4   &  3  & 2 & 1 &   $\frac{1}{2} D_3 D_1^*= \halb \sqrt{n_1 n_3}\, \eulere^{-\imagi\varphi}$   & 0  \\ 
4   &  3  & 3 & 4 &   $-\frac{1}{2} |D_3|^2 =-n_3/2$   & $-g_{4334}$   \\ 
4   &  3  & 4 & 3 &   $\frac{1}{2}  |D_3|^2 =n_3/2$   & $f_{4343}$   \\ \hline\hline
\end{tabular}
\end{table}

\end{appendix}


\begin{thebibliography}{10}
\bibitem{poorresultremark} This is most obvious in the limit of just one NO where $\imagi\partial_t\phi_1=0$ would result. To describe ionization, i.e., an outgoing electron wavepacket, as we study it in Sec.~\ref{sec:autoioni},  a very large set of NO would be necessary. Instead, \reff{eomspaceandNO} captures outgoing wavepackets already with just one NO (and autoionization with two).
\end{thebibliography}
\end{document}